\documentclass[aps,prd,superscriptaddress,showpacs,floatfix,twocolumn,preprintnumbers,nofootinbib]{revtex4}
\usepackage{graphicx}
\usepackage{dcolumn}
\usepackage{amsmath,amssymb,bm}
\usepackage{citesort}

\newcommand{\beq}{\begin{equation}}
\newcommand{\eeq}{\end{equation}}
\newcommand{\bea}{\begin{eqnarray}}
\newcommand{\eea}{\end{eqnarray}}
\newcommand{\no}{\nonumber}

\newcommand{\mpic}{M_{\pi^+}^2}
\newcommand{\mpin}{M_{\pi^0}^2}
\newcommand{\mkc}{M_{K^+}^2}
\newcommand{\mkn}{M_{K^0}^2}
\newcommand{\me}{M_\eta^2}
\newcommand{\mk}{M_K^2}
\newcommand{\mpi}{M_\pi^2}
\newcommand{\fp}{F_\pi}
\newcommand{\fk}{F_K}
\newcommand{\fe}{F_\eta}
\newcommand{\Dpi}{\Delta_\pi}
\newcommand{\DK}{\Delta_K}
\newcommand{\eps}{\epsilon}

\newcommand{\Lag}{{\cal L}}
\newcommand{\M}{{\cal M}}
\newcommand{\Order}{{\cal O}}
\newcommand{\eq}{~=~}
\newcommand{\re}{{\rm Re}\,}
\newcommand{\im}{{\rm Im}\,}

\begin{document}

\preprint{FZJ-IKP-TH-2007-21, HISKP-TH-07/18}

\title{Investigation of \boldmath{$a_0-f_0$} mixing}

\author{Christoph Hanhart}
\email{c.hanhart@fz-juelich.de}
\affiliation{Institut f\"ur Kernphysik (Theorie), 
             Forschungzentrum J\"ulich, D-52425 J\"ulich, Germany.}

\author{Bastian Kubis}
\email{kubis@itkp.uni-bonn.de}
\affiliation{HISKP (Theorie),
             Universit\"at Bonn, Nussallee 14-16, D-53115 Bonn, Germany.}

\author{Jos\'e R.\ Pel\'aez}
\email{jrpelaez@fis.ucm.es}
\affiliation{Departamento de F\'{\i}sica Te\'orica II, 
             Universidad Complutense, E-28040, Madrid, Spain.}

\date{\today}

\begin{abstract}
We investigate the isospin-violating
mixing of the light scalar mesons $a_0(980)$ and $f_0(980)$ 
within the unitarized chiral approach. Isospin-violating
effects are considered to leading order in the quark mass differences and
electromagnetism.  In this approach both mesons are generated through
meson-meson dynamics.  
Our results provide a description of the mixing phenomenon 
within a framework consistent with chiral symmetry
and unitarity, where these resonances are not predominantly $q\bar{q}$ states.
Amongst the possible experimental signals, we discuss
observable consequences for the reaction $J/\Psi\to \phi \pi^0\eta$ in detail.
In particular we demonstrate that the effect of $a_0-f_0$ mixing is
by far the most important isospin-breaking effect in the resonance region
and can indeed be extracted from experiment.
\end{abstract}

\pacs{12.39.Fe, 13.20.Gd, 13.40.Ks}

\maketitle

\section{Introduction}

Although the light scalar mesons $a_0(980)$ and $f_0(980)$
have been established as resonances long ago, there is still a
heated debate going on in the literature regarding the very nature of these states.
Naively one might
assign them a conventional $q\bar q$ structure, however, at
present no quark model is capable of describing both states
simultaneously as $q\bar q$ states --- see, e.g., Ref.~\cite{bonn}. 
On the other hand, as early as 1977 it was stressed that especially in
the scalar channel the interaction of four-quark systems (two quarks, two
antiquarks) is attractive~\cite{jaffe}. Some authors have found indications
 for the existence of compact four-quark
states~\cite{achasov1,vijande}. However, the same short-ranged interaction can
also be the kernel to the scattering of pseudoscalars, giving rise to
extended four-quark states that one might call hadronic molecules or
extraordinary hadrons~\cite{jaffenew1,largeNc1,largeNc2,largeNc3}. Independently, a similar conclusion
was found in different approaches~\cite{isgur1,isgur2,isgur3,janssen,eef,Oller:1997ti}.

A very different approach to quantify the nature of scalar states was
presented in Ref.~\cite{evidence}, where it was argued that the value
of the effective coupling constant of a resonance to a particular
continuum channel is a direct measure of its molecular component if
the corresponding threshold is very close to the resonance
position. When applied to the case of the $f_0(980)$ also this 
model-independent analysis revealed that this scalar is (to a very large
degree) of molecular nature~\cite{evidence}. This picture was further
supported by analyses of the reactions $\phi\to \pi^0\pi^0\gamma$ and
$f_0\to \gamma\gamma$~\cite{yulia,yulia2}.  For the $a_0(980)$, on the other
hand, no clear picture emerged from these studies (note that also
the data is of poorer quality). This might either
mean that the physical $a_0(980)$ has some sizeable admixture of something
different from $K\bar K$ or is a virtual state. Here more information 
is urgently called for.

\begin{figure}
\includegraphics[width=\linewidth]{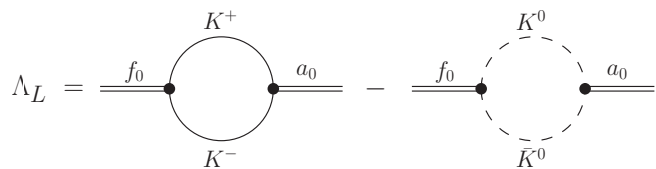}
\caption{\label{lcsb} 
Graphical illustration of the leading contribution to the $a_0-f_0$ mixing
matrix element $\Lambda_L$ defined in Eq.~\eqref{llam}. 
}\end{figure}

An observable supposedly of high sensitivity to the structure of the
scalar mesons was identified at the end of the 1970s, when Achasov and
coworkers observed that the isospin-violating mixing of the isovector
$a_0$ and the isoscalar $f_0$ should be significantly enhanced due to
the proximity of the kaon thresholds to the poles of both mesons. In
Ref.~\cite{achasovmix} it was demonstrated that the leading piece of
the $a_0-f_0$ mixing amplitude can be written as 
\beq \begin{split}
\Lambda_L \,=\, \langle f_0 |T| a_0\rangle 
\,= & ~ i\,g_{f_0K\bar K}\,g_{a_0K\bar K}\,\sqrt{s}\,\bigl( p_{K^0}-p_{K^+} \bigr) \\
&+ \Order\bigl(p_{K^0}^2-p_{K^+}^2\bigr) ~,
\label{llam}
\end{split}\eeq
where $p_{K^{0,+}}$ denotes the modulus of the relative momenta of the 
neutral and charged kaon pairs, respectively, 
and the effective coupling constants are defined through 
$\Gamma_{RK\bar K}=g_{RK\bar K}^2p_K$, $R=a_0,\,f_0$. 
Obviously, this leading contribution is just the difference of
the unitarity cut contributions of the diagrams shown in Fig.~\ref{lcsb} 
and is therefore model independent.
In addition, the signal is proportional to the effective coupling constants
of the scalar mesons that encode the essential structure information, as
outlined above.
As already stressed in Ref.~\cite{achasovmix},
the contribution shown in Eq.~\eqref{llam} is
unusually enhanced between the $K^+ K^-$ and the $K^0 \bar K^0$ thresholds, a
regime of only 8~MeV width. Here it scales as
\beq
\sqrt{\frac{\mkn-\mkc}{\mkn+\mkc}} ~\sim~
\sqrt{\frac{m_d-m_u}{m_s+\hat m}} ~,
\eeq
where $m_u$, $m_d$, and $m_s$ denote the current quark masses of the up, down,
and strange quark, respectively, $\hat m=(m_u+m_d)/2$,
and we neglect electromagnetic effects in the kaon masses for this rather symbolic formula.
This is in contrast to common isospin-violating
effects\footnote{By common isospin-breaking effects
we refer to those effects that occur at the
  Lagrangian level.} which scale as $(m_d-m_u)/m_s$, since they have to
be analytic in the quark masses. It is easy to see that away from the kaon
thresholds $\Lambda$ returns to a value of natural size. 
In the phenomenological calculation of Ref.~\cite{krehl} this effect was confirmed. 
However, also there the different kaon
masses were the only source of isospin violation.

\begin{figure}
\includegraphics[width=\linewidth]{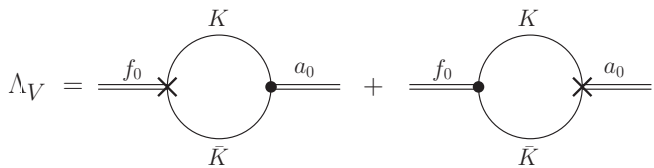}
\caption{\label{nlcsb} 
Graphical illustration of the subleading contribution to the $a_0-f_0$ mixing
matrix element $\Lambda_V$ defined in Eq.~\eqref{nllam}.
The crosses denote isospin-violating vertices. 
}\end{figure}
A subleading  contribution to the mixing amplitude
is given, in the resonance picture, by isospin-violating 
couplings of the resonances to the two-kaon continuum
$g_{RK\bar K}^{\not\hspace{2pt}\rm I}$ as depicted in Fig.~\ref{nlcsb},
\beq \begin{split}
\Lambda_V = 
 i \bigl(g_{f_0K\bar K}^{\not\hspace{2pt}\rm I} \,g_{a_0K\bar K}
+ g_{f_0K\bar K}\,g_{a_0K\bar K}^{\not\hspace{2pt}\rm I} \bigr)
\sqrt{s} \,\,p_K 
+ \Order\bigl(p_K^2\bigr) \,.
\label{nllam}
\end{split}\eeq
Although these effects are regular in the isospin-breaking parameter,
i.e.\ of order $m_d-m_u$, they are kinematically enhanced due to the
unitarity cut $\propto p_K$.  An assessment of the size of such
effects obviously relies on an estimate of the isospin-violating
couplings $g_{RK\bar K}^{\not\hspace{2pt}\rm I}$ that we will provide
in this paper. 
It should be stressed, however, that as their isospin
conserving counterparts, the $g_{RK\bar K}^{\not\hspace{2pt}\rm I}$
are well defined, observable quantities.

\begin{figure}
\includegraphics[width=0.525\linewidth]{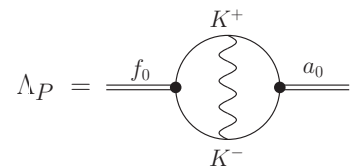}
\caption{\label{photoncsb} 
Soft photon exchange contribution to the $a_0-f_0$ mixing amplitude.
}\end{figure}

In addition there can also be mixing through 
the exchange of soft photons in meson loops --- see Fig.~\ref{photoncsb} ---
giving rise to the mixing amplitude $\Lambda_P$.
The full mixing amplitude is then given by
\beq
\Lambda=\Lambda_L+\Lambda_V+ \Lambda_P \ .
\label{fulllam}
\eeq
If $a_0$ and $f_0$ had a significant admixture
from elementary scalars, one should in addition
expect a direct $a_0-f_0$ transition
to appear. However, this is not included in
our approach.

Our purpose is to improve on the theoretical 
understanding of the possible $a_0-f_0$ mixing phenomenon.
Especially we would like to get a first quantitative understanding
of the possible impact of isospin-violating couplings.
To do so we employ the chiral unitary approach, in which we will now include
isospin-violating effects to leading order in chiral perturbation theory (ChPT)
in both the strong and the
electromagnetic sector. This allows us to address the
following issues:
\begin{enumerate}
\item Does the enhancement of isospin violation in  the effective couplings
  near the $K\bar K$ thresholds, see Eq.~\eqref{nllam}, lead to
  similarly sizeable effects as the kaon mass differences?
\item What is the effect of soft photon exchange on
the mixing amplitude, see Fig.~\ref{photoncsb}?
\item What is the resulting mass dependence of the signal  for
  the mixing?
\item In Refs.~\cite{aipproc,report} it was claimed that, in the mass range considered,
$a_0-f_0$ mixing is the by far dominant isospin-violating effect as it emerges
from
the overlap of two narrow resonances with very nearby masses. 
This issue is discussed in Sec.~\ref{sec:isoFSI}.
As we will see, we are now in
the position to check this estimate within a dynamical approach for a specific
reaction.
\end{enumerate}

For our calculations we use the chiral unitary approach as
developed in Refs.~\cite{Oller:1997ti,Oller:1998hw}, which
provides the  amplitudes for the scattering of two pseudoscalars
from the coupled channel unitarization of the leading-order chiral Lagrangian. 
What is interesting about this approach is that it is not only able to describe 
the data on pseudoscalar-pseudoscalar $S$-wave scattering up to 1.2~GeV remarkably well, 
with just one cutoff of natural size, but 
also to dynamically generate the poles associated with the lightest scalar mesons
without the need to introduce them explicitly in the Lagrangian. 
Hence one can avoid \textit{a priori} assumptions about the nature or 
even the existence of those resonances 
that come out naturally as a consequence of chiral 
symmetry and coupled channel unitarity. 
This reduces  the model dependence of the approach considerably.
In order to establish the nature of the generated poles, 
additional theoretical information is necessary, e.g.\ in Ref.~\cite{largeNc1}
the leading $1/N_c$ behavior was used to provide evidence for a non-$q\bar{q}$ nature of the scalars.
The scattering amplitudes obtained this way have then been used to 
implement the final-state interaction in the next-to-leading-order
calculation of scalar form factors~\cite{MO,timo}. 

Another relevant aspect of unitarized ChPT is that it can be extended 
to higher orders, and, indeed, it is also possible to use 
the fully renormalized next-to-leading-order unitarized ChPT scattering amplitudes
\cite{Dobado:1996ps,Dobado:1992ha,GomezNicola:2001as,Pelaez:2004xp} to match the final-state interactions 
into form factors~\cite{Dobado:1999xb}.  However, for the scalar form factors
we are interested in, it is possible to simplify the approach
by matching their next-to-leading-order calculations 
just with the leading-order chiral unitary scattering amplitudes.
Indeed, it has been shown \cite{MO,timo} that this approximation already
provides a very good 
description of the existing data on isospin-conserving
processes, and thus provides a well founded
starting point for our approach.

In the literature various reactions are discussed that should
be sensitive to the isospin-violating $a_0-f_0$ mixing;
amongst those are $\gamma p\to p\pi^0 \eta$~\cite{tabakin},
$\pi^-p\to\pi^0\eta n$~\cite{achasovPRL,achasovPRD}, $pn\to
d\pi^0\eta$~\cite{sasha1,sasha2,sasha3}, $dd\to \alpha \pi^0\eta$~\cite{leonid}, and
$J/\Psi\to \phi \pi^0\eta$~\cite{close,wu}. In the first three certain
differential observables are sensitive to $a_0-f_0$ mixing, for the
last two the cross section is proportional to the square of the mixing
amplitude, since the corresponding amplitudes vanish in the isospin limit.
In this paper we will focus on the last reaction since the recent
measurement by the BES collaboration of the isospin-conserving channel
$J/\Psi\to \phi \pi^+\pi^-$~\cite{BES} shows a very pronounced signal
of the $f_0$.\footnote{In principle one could also study $J/\Psi\to \omega \pi^0\eta$,
however, there is no clear signal of the $f_0$ visible in the
corresponding isospin-conserving channel $J/\Psi\to \omega \pi \pi$,
and therefore also no pronounced mixing amongst $a_0$ and $f_0$ 
should be expected in this channel.}  In addition these
data were already analyzed within the unitarized chiral approach in
Ref.~\cite{timo}, going back to the formalism developed in
Ref.~\cite{MO}, which is very convenient for our purposes since it
determines the isospin-conserving part of our
formalism rather accurately.  A variant of this analysis, including isospin-breaking
sources, forms the basis of our study.

\section{Theoretical framework for the isospin-conserving case}

\subsection{Isospin states and scalar form factors}

In all reactions listed above, where signals of $a_0-f_0$ mixing
are expected, there are three strongly interacting particles in the
final state. In principle this necessitates a full three-body
treatment using relativistic Faddeev equations. 
However, since we will focus on a phenomenon that occurs within a very narrow kinematic
window, we will adopt the usual approximation that kinematics can be chosen such that
the interactions within the particle pair of interest can be
isolated. This approximation has already been demonstrated to 
provide a good description of the data~\cite{MO,timo},
but should be checked within
a Dalitz plot analysis of the data, once available. 

Therefore, in what follows we assume that only the interaction
of the two outgoing pseudoscalar particles needs to be considered.
The full production amplitude is then given by the scalar
form factor times at most a polynomial~\cite{leutwyler}.
Since we are interested in a  phenomenon that occurs
within a mass range of a few tens of MeV only, we can safely 
use a polynomial of zeroth order only.

The considerations above are rather general and should hold
for all reactions listed above as possible candidates to find
signals of $a_0-f_0$ mixing. 
As a concrete example and since it will be used below,
we now briefly reiterate the formalism of Refs.~\cite{MO,timo} to 
describe the decay of the $J/\Psi$ into a $\phi$ and two pseudoscalars.
We use a Lagrangian coupling the two vector particles
to scalar currents of zero isospin in the form
\beq
\Lag \eq C_\phi \Psi^\mu \phi_\mu \bigl[ \bar ss + \lambda_\phi \bar nn \bigr] ~,
\label{eq:LagrJPsi}
\eeq
where $\bar nn=(\bar uu+\bar dd)/\sqrt{2}$.  The two parameters $C_\phi$, 
$\lambda_\phi$ are \textit{a priori} unknown and have been extracted by fits to experimental
data in Ref.~\cite{MO,timo}.  
Note that these analyses assume $\lambda_\phi$ to be \emph{real},
which amounts to neglecting left-hand cuts or
crossed-channel final-state interactions 
(the latter can in principle be separated in a careful analysis of the Dalitz plot).
The matrix element for the full decay then involves
matrix elements of the scalar currents between the vacuum and two pseudoscalars,
which are described by scalar form factors as follows:
\beq\begin{split}
-\sqrt{2}B\,\Gamma^n_\pi(s) &\eq \langle 0 | \bar nn | \pi\pi\rangle_{I=0} ~, \\
-\sqrt{2}B\,\Gamma^n_K(s) &\eq \langle 0 | \bar nn | K\bar K\rangle_{I=0} ~, \\
-\sqrt{2}B\,\Gamma^n_\eta(s) &\eq \langle 0 | \bar nn | \eta\eta\rangle_{I=0} ~, \\
\end{split}
\label{ffdef} \eeq
and equivalent definitions of the strange scalar form factors with the replacements
$\bar nn \to \bar ss$, $\Gamma^n_i \to \Gamma^s_i$.
The connection of the two-meson states of definite isospin to the basis of physical particles
is given in Appendix~\ref{app:basis}.

\subsection{Form factors and unitarization \label{sec:unit}}

Using ChPT to a certain (e.g.\ next-to-leading) order to calculate the form factors 
defined in Eq.~\eqref{ffdef}
guarantees that we obtain a consistent low-energy expansion,
with the correct chiral loop corrections.  
However, we are interested in 
the energy region of the $a_0$ and $f_0$ resonances, which is outside
the realm of applicability of ChPT amplitudes.  The latter are, 
up to branch cuts generated by Goldstone boson dynamics, 
just polynomials in energy, and as such cannot generate the poles 
that quantum field theory requires to be associated with resonances.
In addition, these polynomials will grow with energy 
and severely violate the unitarity bounds.

It is, however, well known that the two caveats 
above can be fixed by the unitarization of the ChPT 
amplitudes 
\cite{Truong1,Truong2,Truong3,Dobado:1996ps,Dobado:1992ha,GomezNicola:2001as,Oller:1997ti,Oller:1998hw,Pelaez:2004xp}.
In practice, one first unitarizes the partial waves of definite angular momentum $J$
for the scattering of two pseudoscalars, which 
generates the poles associated to the $a_0(980)$ for isospin $I=1$
and $f_0(980)$ for isospin $I=0$ 
(the latter together with the $f_0(600)$, which is of little relevance in the context of this article). 
These unitarized scattering amplitudes are then matched to the form factors
in such a way that the latter have the same poles as the scattering amplitudes, 
and therefore the same resonances.
They do not grow with energy either and
satisfy Watson's theorem of final-state interactions.

Let us briefly review the general formalism before introducing
the necessary modifications for the subsequent inclusion of isospin breaking.
Assuming that only two-body intermediate states are  relevant for the process, 
the unitarity conditions for the $T$-matrix,
once projected onto partial waves of definite angular momentum $J$
(see Appendix~\ref{app:scattamps}), read
\begin{align}
  \im T(s) &\eq T(s)\, \Sigma(s)\, T^*(s)  ~, \label{eq:unitT} \\
  \im \Gamma (s) &\eq T(s)\, \Sigma(s)\, \Gamma^*(s) ~,   \label{eq:unitG}
\end{align}
where
\begin{align} 
T(s) &=
\left(\!\begin{array}{ccc}
T_{11}(s)& T_{12}(s) & T_{13}(s)\\
T_{12}(s)& T_{22}(s) & T_{23}(s)\\
T_{13}(s)& T_{23}(s) & T_{33}(s)
\end{array}\!\right) , ~
\Gamma(s)=\left(\!
  \begin{array}{c}
\Gamma_1(s)\\
\Gamma_2(s)\\
\Gamma_3(s)
\end{array} \!\right), \no \\ \label{eq:vectorgamma}
\Sigma(s) &=\frac{1}{16\pi}
\left(\!\begin{array}{ccc}
\sigma_1(s)& 0 & 0\\
0 & \sigma_2(s) & 0\\
0&0&\sigma_3(s)
\end{array}\!\right) , ~
\end{align}
and $T_{ij}(s)$ is the partial wave $T$-matrix with angular momentum $J$ of the
scattering between states $i$ and $j$. 
The $\sigma_i(s)=2 k_i/\sqrt{s}$ account for the phase space of 
the intermediate two-body states, 
where $k_i$
is the center-of-mass momentum of each physically accessible
state $i$. 

For illustration we have explicitly written the
equations for three coupled states $i,j=1,2,3$, because in the problem at hand,
the $I=0$ and $J=0$ isospin
conserving process corresponds to this formalism with the identification
$1=\pi\pi, 2=K{\bar K}, 3=\eta\eta$, and the $\Gamma_i$ 
are just the form factors defined in Eq.~\eqref{ffdef} above.
Nevertheless, in Refs.~\cite{MO,timo} the calculations were performed in a 
two-channel approach neglecting the final state effects of $\eta\eta$ rescattering,
which will be included here for completeness (and consistency).
Let us remark, finally, that the above formulae as well as
the following results in this section are ready for a straightforward
generalization to the case when the isospin states mix. In particular
we will have to deal with six coupled channels once
we introduce isospin violation in the next sections.

As explained above, the ChPT partial wave $T$-matrices and form factors
cannot satisfy Eqs.~\eqref{eq:unitT} and \eqref{eq:unitG}.
Let us remark that Eq.~\eqref{eq:unitT} implies that 
$T(s)^{-1}= \re T^{-1}(s) -i\, \Sigma(s)$,
so that, indeed, we only have to find an approximation to $\re T^{-1}$. 
Actually, partial waves satisfying the coupled channel unitarity constraint
are obtained from the following expression~\cite{Oller:1997ti,Oller:1998hw}:
\beq
  \label{eq:tunitarized}
  T(s)=\bigl[I-T^{(2)}(s)\, G(s)\bigr]^{-1}\,T^{(2)}(s) ~,
\eeq
where $I$ is the identity matrix, 
$T^{(2)}$ is the leading-order ChPT $T$-matrix, $T=T^{(2)}+\Order(p^4)$, 
and $G(s)$ is a diagonal matrix whose
elements $G_i(s)$ are one-loop integrals corresponding to the two mesons 
of the state $i$ propagating in the loop; detailed expressions are provided in 
Appendix~\ref{app:G}. 
Note that, in the physical (or charge) basis, $G(s)$ is a matrix whose 
diagonal elements $G_i(s)$ are analytic functions except for
a right cut starting at each $i$ state threshold.
The imaginary part of $G_i(s)$ is precisely $\sigma_i(s)/16 \pi$.
Moreover, if $T$ is reexpanded, one recovers the leading-order ChPT result including
the correct imaginary part obtained at one loop. 
In addition, the factor $[I-T^{(2)}\,G(s)]^{-1}$
generates the required poles associated with resonances.
Alternative derivations of this unitarization formalism make use of Lippmann--Schwinger-like
equations \cite{Oller:1997ti} or dispersive approaches \cite{Dobado:1996ps,Dobado:1992ha,GomezNicola:2001as}.

Unitarization thus provides the summation 
of the two-meson $s$-channel loops, since the $G(s)$ functions yield
the correct imaginary part and a cutoff that can be fixed to approximate
the real part of $\re T^{-1}$, effectively absorbing higher-order terms. 
This has been shown to be sufficient to reproduce the available scattering
data on the scalar channel and generates the observed resonances~\cite{Oller:1997ti,Oller:1998hw}. 
Of course, with one natural cutoff this approximation
is not always valid over the whole light resonance regime, and indeed
next-to-leading-order terms are necessary to generate other states, like vectors,
from unitarization~\cite{Dobado:1996ps,Dobado:1992ha,GomezNicola:2001as}.

Equation~\eqref{eq:tunitarized} shows how to unitarize scattering amplitudes, 
but starting from this expression it is straightforward to also
obtain scalar form factors that satisfy Eq.~\eqref{eq:unitG},
writing~\cite{MO}
\begin{equation}
  \label{eq:Gammaunitarized}
  \Gamma(s) = \bigl[I-T^{(2)}(s)\, G(s)\bigr]^{-1}\,R(s) ~,
\end{equation}
where $R(s)$ is a vector of real functions free from any singularity,
which can be determined from a matching to the next-to-leading-order ChPT calculation 
of the form factors $\Gamma=\Gamma^{(0)}+\Gamma^{(2)}+\Order(p^4)$ \cite{Gasser:1984ux}.
By reexpanding $[I-T^{(2)}(s)\, G(s)]^{-1}\simeq I+T^{(2)}(s)\, G(s)+...$, 
one can extract $R(s)$ from
\begin{equation}
  \label{eq:Rdef}
  \Gamma(s) ~=~ \bigl[I+T^{(2)}\,G(s)\bigr]R(s)+ \Order(p^4) ~.
\end{equation}
Since the $G(s)$ integrals do have a residual cutoff $q_{max}$ dependence, 
not present in the
renormalized next-to-leading-order ChPT calculation of $\Gamma$,
this matching has to be performed at a given renormalization scale  $\mu=1.2 q_{max}$
(see \cite{Oller:1998hw} for the relation between the cutoff and dimensional 
regularization of the $G(s)$ functions).
We provide the explicit expressions for $R(s)$ in Appendix~\ref{app:R}.
We will now discuss how this formalism  needs to
be extended to include the effects of isospin violation.

\section{Isospin violation, theoretical framework}

Let us focus on the production reaction
$J/\Psi\to \phi \pi^0\eta$, and calculate the matrix element
\beq
\M_\phi^{\pi\eta} \eq C_\phi \langle 0 | \bar ss+\lambda_\phi \bar nn | \pi^0 \eta \rangle ~.
\label{eq:Mpieta}
\eeq 
In analogy to what we described above for the isospin-conserving case, 
the matrix element will be written in terms of scalar form factors
\beq 
\begin{split}
-\sqrt{2} B \, \Gamma^n_{\pi\eta}(s) &\eq \langle 0 | \bar nn | \pi^0\eta \rangle ~,\\
-\sqrt{2} B \, \Gamma^s_{\pi\eta}(s) &\eq \langle 0 | \bar ss | \pi^0\eta \rangle ~,\\
\end{split} \label{eq:defGpieta}
\eeq
which obviously vanish in the isospin limit, as the source terms are isoscalar,
while the final state has $I=1$.
We will start with a variant of the recent analysis of 
the reaction $J/\Psi\to \phi \pi^+\pi^-$~\cite{timo}. 
There the data reported by the BES collaboration~\cite{BES} was
analyzed using the same unitarized chiral approach, sketched in
the previous section, for the meson-meson final-state interaction.
Thus all that needs to be done to investigate the effect
of isospin violation is to replace the matrix $G(s)$, the vector $R(s)$, 
and the meson-meson scattering
matrix used there by those including isospin violation.

\subsection{Form factor unitarization with isospin violation}

Unitarization does not actually rely on isospin conservation, but is just
a formalism derived for partial waves that could have been applied equally well in 
the physical (charge) instead of the isospin basis,
which is indeed the most appropriate, once we allow for isospin breaking.
Isospin-breaking effects in unitarized chiral effective theories
have been studied before in the context of $\eta$ and $\eta'$ decays, see
Refs.~\cite{Beisert,Borasoy}.

The enhancement of isospin violation we are interested in is due to the
fact that we are looking at the region of the two $K\bar{K}$ thresholds, where
the dynamics is dominated by the resonances already 
generated within the chiral unitary approach. Isospin breaking is a small correction
to the isospin-conserving formalism, which simply amounts to increasing the number of distinct
states and to slightly modifying the structure of the vertices.

In the following sections we will thus perform the calculation
of the different pieces needed in order to include isospin violation
in the chiral unitary approach to the scalar form factors, as described in Sec.~\ref{sec:unit}.
In particular, in addition to the three $I=0$ states,
we now also have to consider two $I=1$ and one $I=2$ states.
Hence, the $T^{(2)}$ matrix is now six-dimensional, and its elements
are easily obtained from the partial waves, shown in Appendices~\ref{app:basis}
and \ref{app:scattamps}. The matrix of loop functions is also six-dimensional,
as seen in Eq.~\eqref{eq:Gsixdim}.

We can  obtain all unitarized form factors 
for $J/\Psi$ decays into a $\phi$ plus two $S$-wave pseudoscalar mesons
in the isospin basis from the following equation:
\begin{widetext}
\begin{equation}
  \label{eq:IVgammavectors}
\left(
    \begin{array}{c}
\Gamma^{I=2}_{\pi\pi}(s)\\[1mm]
\Gamma^{I=0}_{\pi\pi}(s)\\[1mm]
\Gamma^{I=0}_{\eta\eta}(s)\\[1mm]
\Gamma^{I=0}_{K\bar{K}}(s)\\[1mm]
\Gamma^{I=1}_{K\bar{K}}(s)\\[1mm]
\Gamma^{I=1}_{\pi \eta}(s)
    \end{array}
\right)
  ~=~ \bigl[I- T^{(2)}(s)\,G(s)\bigr]^{-1}\left(
    \begin{array}{c}
0\\[1mm]
R^s_{\pi}(s)+\lambda_\phi R^n_{\pi}(s)\\[1mm]
R^s_\eta(s)+\lambda_\phi R^n_{\eta}(s)\\[1mm]
R^s_{K}(s)+\lambda_\phi R^n_{K}(s)\\[1mm]
\tilde \lambda_\phi R^{ud}_K(s)\\[1mm]
R^s_{\pi\eta}(s)+\lambda_\phi R^n_{\pi\eta}(s)+\tilde \lambda_\phi R^{ud}_{\pi \eta}(s)
    \end{array}
\right) ~,
\end{equation}
\end{widetext}
where all the $R$ functions can be found 
in Appendix~\ref{app:R} and in Eqs.~\eqref{eq:Rpieta} and \eqref{eq:Rud}. 
The $I=1$ polynomial form factor terms $\propto \tilde \lambda_\phi$,
as well as the appearance of $I=0$ form factors $R^{n,s}_{\pi\eta}(s)$ 
in the $\pi\eta$ channel 
will be discussed in detail in Sec.~\ref{sec:IVPO}.
We neglect an
isospin-violating production vertex with $I=2$ since it would be subleading
in $m_d-m_u$.
Similarly, contributions $\propto \tilde\lambda_\phi R^{ud}_{\pi,K,\eta}(s)$
could be added to the $I=0$ production terms,
but are neglected as second order in isospin violation.
Let us remark that if we turn off isospin violation in $T^{(2)}$
and $G(s)$ and set $\tilde \lambda_\phi=0$, 
$R^n_{\pi\eta}(s)= R^s_{\pi\eta}(s)=0$, we recover the 
three coupled channel isospin-conserving case.

For the study of $a_0-f_0$ mixing, we are interested in the
$I=1$ form factors and more precisely in $\Gamma^{I=1}_{\pi\eta}(s)$
in the region around $K\bar{K}$ threshold.
Note that Eq.~\eqref{eq:IVgammavectors} is very convenient in order to switch on and off
the different contributions to isospin violation,\
and study their sizes,
since all the isospin violation in  vertices appears in $T^{(2)}$,
and the difference between charged and neutral loops appears in $G(s)$.
This will be studied in Sec.~\ref{sec:results}.

\subsection{Why \boldmath{$a_0-f_0$} mixing should dominate isospin violation}\label{sec:isoFSI}

In Ref.~\cite{aipproc,report} it was claimed that, if the invariant
mass of the outgoing two-meson pair is close to the nominal mass of
both $f_0$ and $a_0$, then the mixing of the two should be, by far, the
dominant isospin-violating effect.  The argument was based on the fact
that the two scalar resonances of interest are narrow and overlap and
therefore the effect of isospin violation as it occurs in the
propagation of the scalar mesons is enhanced compared to mixing in the
production operator. In Refs.~\cite{aipproc,report} the reasoning was
presented for $NN$ induced production of the scalar mesons.  Here we
adopt it for $J/\Psi$ decays. 
For the sake of this argument only, we introduce the notion of explicit
resonance ($a_0$, $f_0$) propagators; we emphasize, though, that
such objects never appear in this form in our unitarized chiral amplitudes,
where the corresponding poles are generated dynamically.

We expect the effect of isospin
violation in the production operator ($\propto \tilde \lambda_\phi$ in
Eq.~\eqref{eq:IVgammavectors}) to scale at most as
$(m_d-m_u)/m_s$. This is then followed by an isospin-conserving 
final-state interaction proportional to $W_{a_0\to\pi\eta}G_{a_0}$, where
$G_{a_0}$ denotes the $a_0$ propagator and $W_{a_0\to \pi\eta}$ the
$a_0$ decay matrix element --- if we assume a scalar coupling of the
$a_0$ to $\pi\eta$, the vertex function $W_{a_0\to \pi\eta}$ reduces
to the effective coupling constant $g_{a_0\pi\eta}$ (c.f.\ the
corresponding couplings to the kaon channels defined in
Eq.~\eqref{llam}). On the other hand, $a_0-f_0$ mixing occurs in the
propagation, here parameterized by the various isospin-violating
scalar form factors $\Gamma_{\pi\eta}^{n,s}$. We may therefore use
\beq
\Gamma_{\pi\eta}^{n,s} ~\sim~ W_{a_0\to \pi\eta} \, G_{a_0} \, \Lambda \, G_{f_0} ~.
\eeq
Here $\Lambda$ denotes the mixing matrix element, see Eq.~\eqref{fulllam}.  
As it has been argued in the introduction, $\Lambda$ scales
as $M_K^2\sqrt{(m_d-m_u)/m_s}$.  
For the $f_0$ propagator very close to the $K\bar K$ threshold,  
we may use 
\beq
G_{f_0} ~=~ \frac{1}{s-m_{f_0}^2+i\,m_{f_0}\Gamma_{f_0}} ~,
\eeq
which reduces to $-i/(m_{f_0}\Gamma_{f_0})$ for $s\simeq m_{a_0}^2\simeq m_{f_0}^2$. 
Thus, the ratio of the two effects, 
and therefore an estimate of the theoretical uncertainty of the investigation, 
is given by $m_{f_0}\Gamma_{f_0}\sqrt{(m_d-m_u)/m_s}/M_K^2$, which is of the order of 4\% for the
amplitude (for this estimate we used 50~MeV for the width).  
This estimate can now be tested within a dynamical approach.

\subsection{Lagrangians, mass splittings etc.}

We use the leading order chiral Lagrangian $\Lag^{(2)}$ including
isospin-breaking/electromagnetic effects~\cite{urech}:
\beq
\Lag^{(2)} = \frac{F^2}{4} \langle D_\mu U D^\mu U^\dagger
+ \chi U^\dagger + U \chi^\dagger \rangle
+ C \langle QUQU^\dagger \rangle ~.\label{L2}
\eeq
$U$ collects the Goldstone boson fields in the usual manner,
$F$ is the common pseudoscalar meson decay constant.
The covariant derivative in particular contains the coupling to photons,
$D_\mu U = \partial_\mu U - iQ[A_\mu,U] + \ldots\,$,
and $Q$ is the quark charge matrix, $Q=e\,{\rm diag}(2,-1,-1)/3$.
The field $\chi$ collects the quark masses, $\chi = 2B\,{\rm diag}(m_u,m_d,m_s)+\ldots\,$.
We choose to express isospin-breaking effects due 
to the light quark mass difference in terms of the 
leading order $\pi^0\eta$ mixing angle
\beq
\eps \eq
\frac{\sqrt{3}}{4}\frac{m_d-m_u}{m_s-\hat{m}} + \Order\bigl((m_d-m_u)^3\bigr) ~.
\eeq
Electromagnetic contributions $\propto C$ in Eq.~\eqref{L2}
can be reexpressed in terms of the charged-to-neutral pion mass difference,
\beq
\Dpi = \mpic - \mpin = \frac{2Ce^2}{F^2} ~,
\eeq
where the tiny strong mass difference $\propto (m_d-m_u)^2$ is neglected.
Because of Dashen's theorem, the charged-to-neutral kaon mass difference can 
then be written at leading order as
\beq
\DK = \mkc - \mkn = \Dpi - \frac{4\eps}{\sqrt{3}} \left(\mk-\mpi\right) ~.
\eeq

As outlined in the introduction, the meson mass differences,
especially those of the kaons, naturally introduce a striking
isospin-violating effect.
Note that, as a consequence of the previous arguments,
the matrix of loop functions $G(s)$ is now six-dimensional,
i.e.\ is  diagonal in the charge basis
\begin{equation}
G={\rm diag}\bigl(G_{\pi^+\pi^-},G_{\pi^0\pi^0},G_{\eta\eta},G_{K^+K^-},G_{K^0\bar K^0},G_{\pi^0\eta}\bigr)\label{eq:Gsixdim} ~,
\end{equation}
but has only a block diagonal form in isospin basis (see Appendix~\ref{app:basis}).

We have pointed out in Eq.~\eqref{llam} that the mass difference in 
the kaon propagators generates the leading isospin-breaking contribution
in the $a_0$/$f_0$ resonance region.  
As we are about to calculate subleading isospin-violating effects, 
one may wonder how accurately the unitarity cut contribution 
is described by the (leading-order) chiral unitary approach, 
and how much this description would change if we considered
unitarized chiral $p^4$ amplitudes (see Refs.~\cite{Dobado:1996ps,Dobado:1992ha,GomezNicola:2001as}).
In this formalism we do not introduce  the couplings of the $a_0$ and $f_0$ 
to different isospin channels explicitly, which, as seen in Eq.~\eqref{llam},
determine the strength of the leading part of the mixing amplitude. 
Actually, within this approach, those couplings correspond to the 
residues of the poles that have been generated dynamically when fitting the data.  
In this sense the couplings are determined by the set of data that has been 
fitted to obtain the isospin-conserving part.  
The formalism as presented is very general and for the $S$-waves, 
the quality of such fits depends mainly on the data considered, 
and hardly changes with the order of the unitarized chiral amplitudes. 
Of course, whenever new data appears for the isospin-conserving reactions, 
the corresponding parts can be refitted and therefore allow for 
an improvement in the accuracy also for the isospin-violating amplitudes.  
However, the relevant observation is that at any time the possible effect 
of isospin violation in the couplings, as discussed in the following section,
should be considered.

\subsection{Isospin violation in vertices}

Isospin violation in the Lagrangian in Eq.~\eqref{L2} does not only induce charged-to-neutral
pion and kaon mass differences, but also affects the (tree-level)
scattering amplitudes, which receive corrections to their 
isospin-conserving expressions.  This formalism allows one to calculate isospin violation
in scattering consistently with the analysis of the masses.
A well-known example for the importance of such effects
is the sizeable correction in relating $\pi\pi$ scattering amplitudes
at threshold to the isospin scattering lengths~\cite{KnechtUrech};
similar corrections have also been calculated e.g.\ in $\pi K$ 
scattering~\cite{Kubis1,Nehme1,Nehme2,Kubis2}.
In addition, some transitions only take place at all in the presence of isospin 
violation, the most prominent being the decay $\eta \to 3\pi$~\cite{Osborn,GL85eta}.

The isospin-violating 
scattering $T$-matrix has to be calculated in the particle or charge basis;
the complete list of amplitudes, linking all charge- and strangeness-neutral 
channels, calculated at 
leading order within ChPT, i.e., $T\simeq T^{(2)}+\Order(p^4)$,
is given in Appendix~\ref{app:scattamps}.
Note that we now have six coupled states, but 
as we have commented in Sec.~\ref{sec:unit}, the same unitarization formalism applies.
Despite obtaining our calculations in the charge basis,
it is still convenient to recast them in
the isospin basis, as we assume (for the moment)
production of a pure $I=0$ state, while $\pi^0\eta$ is $I=1$. We show its relation
to the charge basis in Appendix~\ref{app:basis}.
Obviously, the matrix of scattering amplitudes in the isospin basis is not
block diagonal with respect to $I=0,\,1,\,2$ anymore, but allows for transitions
between different isospin quantum numbers; these transition matrix elements
scale with either $\epsilon$ or $e^2$.

In the resonance picture of Eqs.~\eqref{llam} and \eqref{nllam},
isospin violation in the meson-meson vertices induces the isospin-violating
resonance couplings $g_{RK\bar K}^{\not\hspace{2pt}\rm I}$.
As we have no strict counting scheme for energies in the resonance region,
these resonance couplings are only modeled this way,
and may receive corrections from higher orders, although one should take into account that
unitarization is necessarily taming their effect.
They are not fixed by isospin-symmetric data in the way the 
cut contribution due to kaon mass differences are,
and therefore have to be pinned down directly from isospin-violating decays,
like the one discussed here.

\subsection{Coulomb corrections}\label{sec:Coulomb}

\begin{figure}
\includegraphics[width=0.85\linewidth]{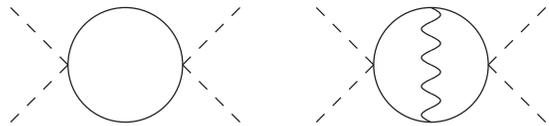}
\caption{\label{fig:photonloop}
Photon exchange diagram that generates a singular
behavior at the $K^+K^-$ threshold (right).
The full lines denote charged kaons, dashed lines arbitrary 
other mesons, and the wiggly line the exchanged photon. 
This is added to the standard $K^+K^-$ one-loop function (left)
in the iterated bubble sum.
}\end{figure}

So far, all interaction terms derived from Eq.~\eqref{L2},
entering the matrix $T^{(2)}$ in the unitarized final-state interaction,
are pointlike four-meson vertices.  
We note, however, that Eq.~\eqref{L2} generates another 
tree-level diagram contributing to meson-meson scattering,
namely one-photon exchange between charged mesons.
It is obvious that this nonlocal interaction cannot be taken into account
on quite the same footing.  

As all initial- and final-state particles in the decay $J/\Psi \to \phi \pi^0 \eta$
are electrically neutral, photons can only enter inside charged-meson loops.
The diagram shown in Fig.~\ref{fig:photonloop} is the only one at $\Order(\alpha)$
that is enhanced at the $K^+K^-$ threshold;  
we neglect all other, nonenhanced diagrams.
Our prescription is then to replace the charged-kaon loop function $G_{K^+K^-}(s)$
in the unitarization sum by the sum of this and the one-photon exchange graph,
\beq
G_{K^+K^-}(s) \to G_{K^+K^-}(s) + G^{1-C}_{K^+K^-}(s) ~.
\eeq
In the threshold region, the exact expression for the one-photon-exchange diagram~\cite{Broadhurst} 
can well be approximated by the threshold-expanded form~\cite{BenekeSmirnov},
which reads
\beq \begin{split}
G^{1-C}_{K^+K^-}(s) &= 
\frac{\alpha}{32\pi} \biggl\{
\log\frac{|s-4\mkc|}{\mkc} + \log 2 + \frac{21\zeta(3)}{2\pi^2} \\
& - i\,\pi\,\theta\bigl(s-4\mkc\bigr)\biggr\} + \Order\bigl((s-4\mkc\bigr)^{1/2}\bigr) ~.
\end{split} \eeq
A justification for this being the leading threshold behavior can also 
be given in the framework of a nonrelativistic theory~\cite{pionium}.

We neglect exchange of multiple photons inside the meson bubble, although
these are in principle also enhanced close to threshold.  A resummation of 
multiphoton exchange is necessary once the parameter
$$ \frac{\alpha}{2} \biggl(1-\frac{4\mkc}{s}\biggr)^{-1/2} $$
is not small anymore;
it, e.g., becomes as large as 0.1 for $\sqrt{s}= 2M_{K^+} \pm 0.7$~MeV.
The size of the more-than-one-photon exchange amplitude relative to one-photon one
stays below 5\% outside a window of $\pm 5$~MeV around the charged kaon threshold.
As the energy resolution for potential experiments is expected to 
be of this order at best, neglecting higher-order photon graphs seems well justified.
This is in line with the findings of Ref.~\cite{kkwithphoton}.

Note that we only take photon exchange into account for the charged \emph{kaon} 
loop graphs.  We have checked that the corresponding modification inside
the charged pion loops leads to no visible modification in the physical region
of the process under investigation,
i.e.\ above $\pi^0\eta$ threshold, and in particular not in the 
energy region around $K\bar K$ threshold considered here.

\subsection{Isospin-violating production operators}\label{sec:IVPO}

It has been argued in Sec.~\ref{sec:isoFSI} that isospin violation in the 
final-state interaction ought to be the dominant effect for the production
of the $\pi^0\eta$ final state in the energy region around the $K\bar K$ threshold.
We can check this assumption explicitly by allowing for isospin breaking
in the production operator.  
This occurs in two forms:
due to mixing, the $\pi^0\eta$ final state has an $I=0$ component,
i.e.\ nonvanishing form factors with the $\bar nn$ and $\bar ss$ currents
that scale with the mixing angle $\epsilon$ exist already at tree level;
and we may allow for an additional $I=1$ component in the scalar source terms
given in Eq.~\eqref{eq:LagrJPsi}.

\subsubsection{Isospin-violating scalar form factors at tree level}\label{sec:IVSFF}

The mixing of $\pi^0$ and $\eta$ leads to nonvanishing form factors
$\Gamma^{n,s}_{\pi\eta}(s)$ already at tree level.
At leading order in the chiral expansion, 
where the propagators for $\pi^0$ and $\eta$ can be diagonalized
with a single mixing angle $\epsilon$, the matrix element 
$\langle 0 | \bar uu+\bar dd+\bar ss | \pi^0\eta\rangle$ has to vanish, 
leading to the relation
$\sqrt{2}\Gamma^n_{\pi\eta}(s) + \Gamma^s_{\pi\eta}(s) = 0$.  
We find
\beq
R^n_{\pi\eta}(s) \eq -\frac{1}{\sqrt{2}} R^s_{\pi\eta}(s) \eq
\frac{2}{3}\epsilon + \Order(\epsilon^3) ~, \label{eq:Rpieta}
\eeq
as $\Gamma(s) = R(s)+\Order(p^2)$, see Eq.~\eqref{eq:Rdef}.
These components are added to the vector of production operators $R$
of Eq.~(\ref{eq:IVgammavectors}).
In contrast to the $I=1$ scalar source term discussed below,
this effect does not induce any additional uncertainty, but comes with
a fixed coefficient.  

In principle, the task of this investigation consists precisely in the determination
of the form factors $\Gamma^n_{\pi\eta}(s)$, $\Gamma^s_{\pi\eta}(s)$, and
$\Gamma^{ud}_{\pi\eta}(s)$.  In contrast to what was done in the isospin-conserving 
case, however, we only match the ChPT expressions for these form factors
at leading order.

\subsubsection{$I=1$ scalar source term}\label{sec:IVSS}

\begin{figure}
\includegraphics[height=3cm]{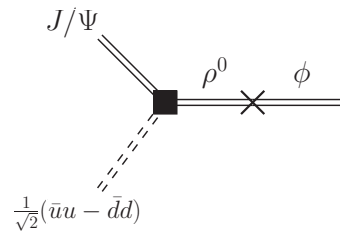}
\caption{\label{fig:mixing}
Vector-meson mixing contribution to $\tilde\lambda_\phi$.  
The box denotes an (isospin-conserving) coupling of $J/\Psi$ to $\rho^0$ 
and the $I=1$ scalar source term, the cross the isospin-violating
$\rho^0-\phi$ mixing term.
}\end{figure}

We first introduce an
$I=1$ scalar source, generalizing the Lagrangian in Eq.~\eqref{eq:LagrJPsi} to
\beq
\Lag \eq C_\phi \Psi^\mu \phi_\mu \Bigl[ \bar ss + \lambda_\phi \bar nn + 
\frac{\tilde\lambda_\phi}{\sqrt{2}}\bigl(\bar uu-\bar dd\bigr) \Bigr] ~.
\eeq
In order to obtain an order-of-magnitude estimate of the size of the new parameter $\tilde\lambda_\phi$,
we invoke a version of vector-meson dominance:  we assume the dominant (isospin-conserving)
production of a $\rho^0$ in association with the operator $(\bar uu-\bar dd)/\sqrt{2}$,
with subsequent $\rho^0-\phi$ mixing; see Fig.~\ref{fig:mixing}.
Following Ref.~\cite{timo}, we write the interaction Lagrangian in the form
\beq
\Lag \eq g \Psi^\mu \Bigl\{ \langle V_\mu \tilde S \rangle 
+ \frac{\nu}{3} \langle V_\mu \rangle \langle S \rangle \Bigr\} ~, \label{eq:LagrVS}
\eeq
where $V_\mu$ collects the vector meson fields, $S$ is the matrix of scalar sources,
and $\tilde S = S - \langle S \rangle/3$.
Considering just the flavor-neutral vector mesons, we can rewrite Eq.~\eqref{eq:LagrVS} as
\beq \begin{split}
\Lag \,=\, \Psi^\mu \Bigl\{ & C_\phi   \,\phi  _\mu \bigl( \bar ss + \lambda_\phi   \bar nn\bigr)
                          + C_\omega\,\omega_\mu \bigl( \bar ss + \lambda_\omega \bar nn\bigr) \\
                         &+ C_\rho  \,\rho^0_\mu \frac{1}{\sqrt{2}}\bigl(\bar uu-\bar dd\bigr) 
\Bigr\}~,
\end{split}\eeq
where $C_\omega$, $\lambda_\omega$, $C_\rho$ can be expressed in terms of $C_\phi$, $\lambda_\phi$ 
according to
\beq
C_\omega = \lambda_\phi C_\phi ~,~~
\lambda_\omega = \frac{1}{\sqrt{2}} + \frac{1}{\lambda_\phi} ~,~~
C_\rho = \Bigl( 1-\frac{\lambda_\phi}{\sqrt{2}}\Bigr) C_\phi ~.
\eeq
$\rho^0-\phi$ mixing is actually assumed to proceed via subsequent
$\rho^0-\omega$ and $\omega-\phi$ mixing.
Neglecting the finite widths of the vector mesons, we find a coupling of the $\phi$ 
to the $(\bar uu-\bar dd)/\sqrt{2}$ operator 
in terms of the $\rho^0-\omega$ and $\omega-\phi$ mixing angles $\Theta_{\rho\omega}$, 
$\Theta_{\omega\phi}$ of the form
\beq
C_\phi \tilde\lambda_\phi \eq C_\rho 
\frac{\Theta_{\rho\omega}\Theta_{\omega\phi}}{(M_\phi^2-M_\rho^2)(M_\phi^2-M_\omega^2)}
~,
\eeq
and therefore
\beq
\tilde \lambda_\phi \eq  \frac{\Theta_{\rho\omega}\Theta_{\omega\phi}}{(M_\phi^2-M_\rho^2)(M_\phi^2-M_\omega^2)}
\Bigl( 1 - \frac{\lambda_\phi}{\sqrt{2}} \Bigr) ~.
\eeq
Plugging in the central values for the mixing angles from Ref.~\cite{ayse}, 
$\Theta_{\rho\omega} = -3.75\times 10^{-3}$~GeV$^2$, $\Theta_{\omega\phi} = 25.34\times 10^{-3}$~GeV$^2$, 
we obtain
\beq
\tilde \lambda_\phi ~\approx~ -0.5 \times 10^{-3} ~, \label{eq:Nlambdatilde}
\eeq
where the uncertainty due to errors in the input numbers is about 20\%.
It is interesting to note that a corresponding estimate based on $\omega-\phi$ mixing
of the $\lambda_\phi$ parameter, which violates the
Okubo--Zweig--Iizuka rule, would lead to
\beq
\lambda_\phi \eq \frac{\Theta_{\omega\phi}}{M_\phi^2-M_\omega^2}
~\approx~ 0.06 ~,
\eeq
which is only about a factor of 2 smaller than the fit result~\cite{timo}.
We therefore assume that the above order-of-magnitude estimate
of $\tilde\lambda_\phi$ should be comparably accurate, and consider it a very conservative
estimate to vary the strength of the isospin-violating production operator within a range
for $\tilde\lambda_\phi$ increased by a factor 
of $\pm 10$ compared to Eq.~\eqref{eq:Nlambdatilde}. 

In order to follow the formalism used earlier, 
we generalize the matrix element Eq.~\eqref{eq:Mpieta} to 
\beq
\M_\phi^{\pi\eta} \eq C_\phi \langle 0 | \bar ss+\lambda_\phi \bar nn 
+\frac{\tilde\lambda_\phi}{\sqrt{2}}(\bar uu-\bar dd) | \pi^0 \eta \rangle ~,
\label{eq:Mpietamod}
\eeq
and define the additional $I=1$ scalar form factors $\Gamma^{ud}_K$, $\Gamma^{ud}_{\pi\eta}$
according to
\beq\begin{split}
-\sqrt{2}B\,\Gamma^{ud}_K(s) &\eq \langle 0 | \frac{1}{\sqrt{2}}(\bar uu-\bar dd) | K\bar K\rangle_{I=1} ~, \\
-\sqrt{2}B\,\Gamma^{ud}_{\pi\eta}(s) &\eq \langle 0 | \frac{1}{\sqrt{2}}(\bar uu-\bar dd) | \pi^0\eta\rangle_{I=1} ~. \\ 
\end{split}\label{eq:defGud}\eeq
As we are only interested in an order-of-magnitude estimate for the effects
of the $I=1$ production operator, we refrain from doing the complete one-loop 
calculation of these form factors and only unitarize the lowest-order ($\Order(p^2)$)
results for these, which we find to be
\beq
R^{ud}_K(s) = \frac{1}{\sqrt{2}} ~, \quad
R^{ud}_{\pi\eta}(s) = -\frac{1}{\sqrt{3}} + \Order(\epsilon^2)~. \quad
\label{eq:Rud}
\eeq
We note that the Feynman--Hellman theorem implies the relation
\beq\begin{split}
R^n_{\pi\eta}(s) &= -\frac{1}{\sqrt{2}} R^s_{\pi\eta}(s) =
-\frac{m_d-m_u}{2(m_s-\hat m)} \,R^{ud}_{\pi\eta}(s) \\
&= \frac{2}{3}\epsilon + \Order(\epsilon^3)
\end{split}\eeq
at tree level, which can easily be checked to be fulfilled
by Eqs.~\eqref{eq:Rpieta} and \eqref{eq:Rud}.

\subsection{Further possible background terms}

In Ref.~\cite{wu}, two candidates for background terms
to an $a_0-f_0$ mixing description of the decay $J/\Psi\to \phi \pi^0 \eta$
were estimated, namely $J/\Psi\to \gamma^* \to \phi \pi^0 \eta$
and  $J/\Psi\to (K^*\bar K+ h.c.)\to \phi\pi^0\eta$.
The former is clearly not covered by our description
of this decay in terms of scalar form factors
and would necessitate a generalization of the production mechanism
beyond $\bar qq$ operators;
luckily it was found to be much smaller than the signal and therefore
will not be considered any further. 

The latter, however, turned out to
be of the order of or even larger than the signal and will now
be discussed briefly.
In the unitarized chiral approach, only the lightest pseudoscalar
mesons appear as dynamical degrees of freedom. The effects
of all other mesons, e.g.\ the $K^*$, are integrated out
and are considered through local counter\-terms.
The corresponding expressions can be deduced from those 
with dynamical heavy particles by formally taking the 
infinite mass limit.

According to Ref.~\cite{wu} the transition in the decay chain
$J/\Psi\to (K^*\bar K+ h.c.)\to \phi a_0$ appears through a triangle
loop that contains two kaon propagators and one $K^*$ propagator;  
isospin violation then emerges through the kaon mass differences. 
In the effective theory description the vector meson propagator needs to be replaced by 
a point interaction --- as a result the whole transition is to 
be regarded as part of the isospin-violating scalar form factor
and therefore, from our point of view, as part of the signal. In
this sense it appears natural that the corresponding transition rate
is of the order of the estimated signal.

\section{Results and discussions}\label{sec:results}

The differential decay rate for the process $J/\Psi \to \phi \pi^0 \eta$ 
is related to the absolute square of the matrix element 
\beq 
\M_\phi^{\pi\eta} = -\sqrt{2}B\,C_\phi \left[
\Gamma_{\pi\eta}^s(s) + \lambda_\phi \Gamma_{\pi\eta}^n(s)  
+ \tilde \lambda_\phi \Gamma_{\pi\eta}^{ud}(s) \right]
~, \label{eq:MFF}
\eeq
with the form factors 
$\Gamma_{\pi\eta}^s(s)$
$\Gamma_{\pi\eta}^s(s) $, and
$\Gamma_{\pi\eta}^{ud}(s)$
given by the matrix relation Eq.~\eqref{eq:IVgammavectors},
by
\beq
\frac{d\Gamma}{d\sqrt{s}} = \frac{\sqrt{
\lambda\bigl(M_{J/\Psi}^2,s,M_\phi^2\bigr)\,
\lambda\bigl(s,\mpin,\me\bigr)}}
{16\sqrt{s}M_{J/\Psi}^2 (2\pi)^3}
\,F_{\rm pol}\, \bigl|\M_\phi^{\pi\eta}\bigr|^2 ~,
\eeq
where $\lambda(x,y,z)=x^2+y^2+z^2-2(xy+xz+yz)$ is the usual K\"all\'en function,
and $F_{\rm pol}$ is the kinematical factor that takes the average and sum
over polarization states of $J/\Psi$ and $\phi$ into account,
\beq \begin{split}
F_{\rm pol}& \eq \frac{1}{3}\sum_{\rho,\rho'} \epsilon_\mu(\rho) \epsilon^\mu(\rho')
\epsilon_\nu^*(\rho)\epsilon^{\nu*}(\rho') \\
& \eq \frac{2}{3} \biggl[ 1+\frac{(M_{J/\Psi}^2+M_\phi^2-s)^2}{8M_{J/\Psi}^2M_\phi^2}\biggr] ~.
\end{split} \eeq

The data for $J/\Psi \to \phi\pi^+\pi^-$ \cite{BES} analyzed in Ref.~\cite{timo}
do not provide normalized differential decay widths, but only event distributions,
$dN/d\sqrt{s} \propto d\Gamma/d\sqrt{s}$, with an unknown constant of proportionality.
Accordingly, we cannot predict a normalized differential decay width for 
$J/\Psi \to \phi\pi^0\eta$ either, but only a \emph{relative} width, 
normalized by the isospin-conserving 2-pion decay channel.
We choose to perform the normalization according to
\beq \begin{split}
& \frac{dN}{d\sqrt{s}} \left(J/\Psi \to \phi\pi^0\eta\right)\biggr|_{\rm norm} \eq  \\
& \frac{d\Gamma}{d\sqrt{s}} \left(J/\Psi \to \phi\pi^0\eta\right) \!\biggl/\! 
\int_{W_1}^{W_2} d\sqrt{s} \frac{d\Gamma}{d\sqrt{s}}\left(J/\Psi \to \phi\pi^+\pi^-\right) \,,
\label{eq:diffnorm}
\end{split} \eeq
where the energy range $W_{1,2}=2\mkc \mp 25$~MeV covers the peak region of
the $a_0-f_0$ mixing signal.
The numerical input on masses and coupling constants that enter our calculations
are given in Appendix~\ref{app:numbers}.
We remark that the coupling constant $C_\phi$
as well as the form factor normalization constant $B$,
see Eq.~\eqref{eq:MFF}, cancel out
in the ratio in Eq.~\eqref{eq:diffnorm} and do not have 
to be specified in our analysis.  
We wish to emphasize that there are \emph{no new} fit parameters
in this analysis:  the unitarization procedure in the chiral unitary
approach at this order contains one single parameter, the cutoff $q_{\max}$,
which has been adjusted as to reproduce the $S$-wave meson-meson scattering data
in all physical channels
(i.e., in particular the resonance pole positions) as well as possible.
All further parameters, both the parameter $\lambda_\phi$ in Eq.~\eqref{eq:LagrJPsi}
as well as the $\Order(p^4)$ chiral low-energy constants $L_i^r$
in the various scalar form factors, 
only serve as a parameterization to describe the isospin-conserving
$J/\Psi \to \phi \pi\pi$ data.

Of course, in order to predict realistic relative 
count rates, one would have to take into account different detector efficiencies
for the two different final states, which we do not do here; it will be
straightforward to implement them, once such experimental specifications become available.

\begin{figure}
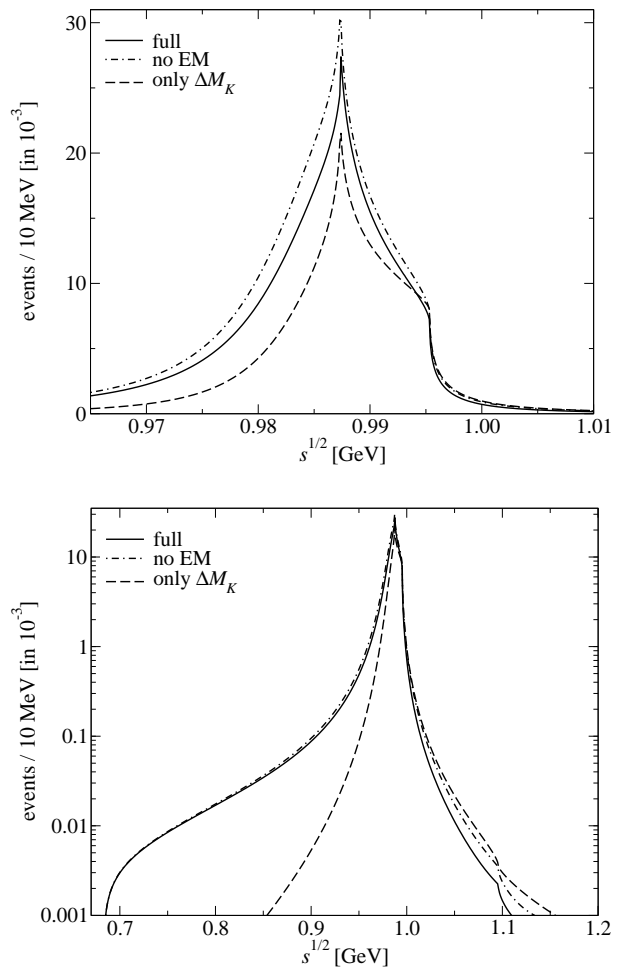

\includegraphics[width=0.925\linewidth]{CountRatepieta_bw}
\vskip 5mm
\includegraphics[width=0.925\linewidth]{CountRatepietawide_bw}
\caption{\label{fig:events}
Predictions for the normalized $J/\Psi \to \phi\pi^0\eta$ differential count rate per 10~MeV,
relative to 1000 $J/\Psi \to \phi\pi^+\pi^-$ events in the peak region
$2\mkc \pm 25$~MeV as defined in Eq.~\eqref{eq:diffnorm}.
We show the successive inclusion of the different isospin-violating effects
in the final-state interaction. 
The kaon mass difference alone leads to the dashed line, 
additional inclusion of isospin violation in the strong vertices
produces the dot-dashed curve.
As the full result, adding one-photon-exchanges in the bubble sum, we obtain the solid curve.
The top panel shows the most relevant region near the two $K\bar K$ thresholds;
in order to make the enhancement in this region more obvious, the
bottom panel shows the whole energy range from threshold up to 1.2~GeV on 
a logarithmic scale.
}\end{figure}
As stated above, the $\pi^0\eta$ invariant mass distribution 
of the reaction $J/\Psi\to \phi \pi^0\eta$ is
proportional to the absolute value squared of the mixing matrix
element, at least as long as we do not include any isospin 
violation in the production operator. In Fig.~\ref{fig:events}
we show our prediction for the $\pi^0\eta$ invariant mass distribution
when different isospin-violating effects are included in the 
propagation of the scalar mesons. To produce the dashed line
only the kaon mass difference was included. The resulting curve 
is in qualitative agreement with those of Refs.~\cite{achasovmix,krehl}.
Including isospin violation in the vertices, we find the dot-dashed curve;  
adding one-photon exchange according to Sec.~\ref{sec:Coulomb}
leads to the full result, given by the solid line.
We find that the effect of photon-exchange is rather small
even in the threshold region, and certainly smaller than the modifications
due to isospin violation in the meson-meson scattering vertices (for
a more detail study of photon effects see Ref.~\cite{kkwithphoton}).
The signal for $a_0-f_0$ mixing is therefore significantly enhanced compared
to the original estimate given in Ref.~\cite{achasovmix}.

\begin{figure}
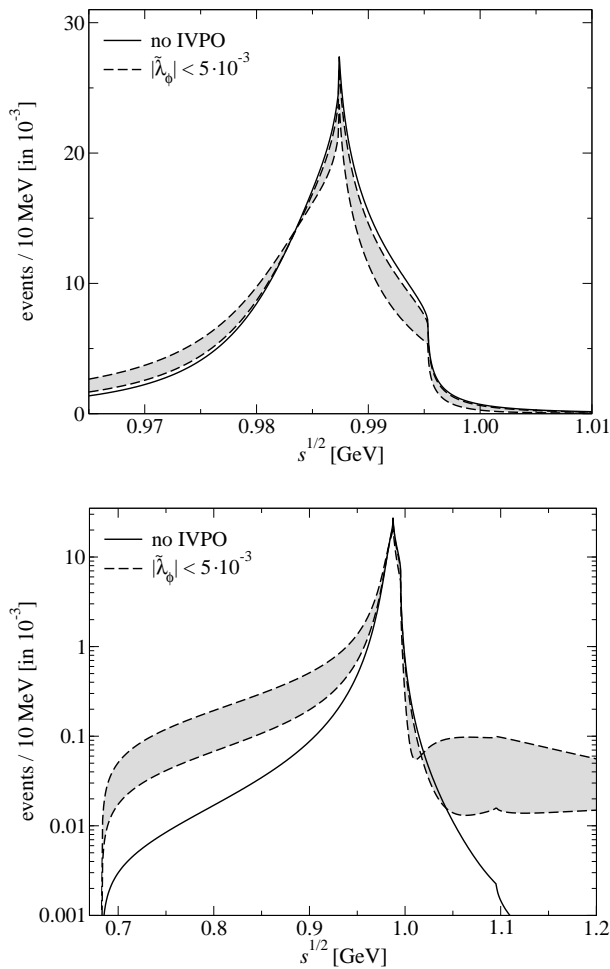

\includegraphics[width=0.925\linewidth]{CountRatepietaB_bw}
\vskip 5mm
\includegraphics[width=0.925\linewidth]{CountRatepietaBwide_bw}
\caption{\label{fig:isoProd}
Predictions for the $J/\Psi \to \phi\pi^0\eta$ differential count rate per 10~MeV
(normalization as in Fig.~\ref{fig:events}),
with (band) and without (full line) inclusion of an isospin-violating production operator (IVPO).
The central curve is shifted by the isospin violation in the scalar form factors 
as discussed in Sec.~\ref{sec:IVSFF},
the band is due to the $I=1$ scalar source with strength limited by
$\tilde\lambda_\phi = \pm 5\times 10^{-3}$, see Sec.~\ref{sec:IVSS}.
As in Fig.~\ref{fig:events}, the top panel shows the two-kaon threshold region,
the bottom one the whole energy range from threshold up to 1.2~GeV on a logarithmic scale.
}\end{figure}

Next we investigate the possible effect of isospin violation in the
production operator. 
As one can see in Fig.~\ref{fig:isoProd}, in the region around the kaon threshold the 
invariant mass distribution in the $\pi^0 \eta$ channel is
by far dominated by the isospin violation in the propagation.
The admixture of an isospin-violating production operator as estimated
in Sec.~\ref{sec:IVPO} actually produces only a rather narrow band
very close to the result without such an operator.
This nicely confirms the corresponding estimates
provided in Ref.~\cite{aipproc,report}.

\begin{table}
\caption{
Event estimates for $J/\Psi \to \phi\pi^0\eta$ for different energy
regions and different isospin-breaking effects.
The energy regions are:
Region I: $M_{\pi^0}+M_\eta \leq \sqrt{s} \leq 2M_{K^+}-25~{\rm MeV}$;
Region II: $2M_{K^+}-25~{\rm MeV} \leq \sqrt{s} \leq 2M_{K^+}+25~{\rm MeV}$;
Region III: $2M_{K^+}+25~{\rm MeV} \leq \sqrt{s} \leq 1.2~{\rm GeV}$.
``$\Delta M_K$'' refers to the original estimate~\cite{achasovmix},
assuming the kaon mass difference as the only source of $a_0-f_0$ mixing.
``Full FSI'' labels the model with all isospin-breaking effects in the
final-state interaction (FSI) included, while ``Including IVPO''
also incorporates the isospin-violating production operators discussed in
Sec.~\ref{sec:IVPO}.  
All event numbers are relative to 1000 $J/\Psi \to \phi\pi^+\pi^-$ events
in Region II. 
\label{tab:events}}
\vskip 3mm
\renewcommand{\arraystretch}{1.5}
\begin{ruledtabular}
\begin{tabular}{cccc}
& Region I & Region II & Region III \\
\hline
$\Delta M_K$   &  0.3                &  20.3                & 0.7                 \\
Full FSI       &  2.4                &  29.0                & 0.4                 \\
Including IVPO & $7.2^{+3.3}_{-2.7}$ & $28.2^{+0.6}_{-0.2}$ & $0.6^{+1.0}_{-0.1}$ \\
\end{tabular}
\end{ruledtabular}
\renewcommand{\arraystretch}{1.0}
\end{table}
In order to demonstrate the enhancement of isospin-breaking in the two-kaon 
threshold region, as well as due to different mechanisms, 
we show the predicted event numbers in Table~\ref{tab:events},
normalized to 1000 $J/\Psi \to \phi\pi^+\pi^-$ events in the peak region,
for three distinct kinematical regions: for the two-kaon 
threshold region of total width 50~MeV, below down to $\pi^0\eta$ threshold, and
above up to 1.2~GeV.  
We want to point out that the signal around 1~GeV is enhanced by nearly 50\%
compared to the original estimate based on the kaon mass difference effect alone.
The reduction due to photon exchange visible in Fig.~\ref{fig:events}
is about 10\%, the corresponding event number
in Region~II of Table~\ref{tab:events} without photon exchange would be 33.6.
Obviously, all mechanisms other than the kaon mass difference also lead to a 
large relative increase of mixing \emph{outside} this central region, in particular
the isospin-violating production operator; this is also clearly seen in the
logarithmic plots over a wider energy range in Figs.~\ref{fig:events} and \ref{fig:isoProd}.
However, a relative enhancement of the signal by more than two orders of magnitude
close to the two-kaon thresholds remains.
Finally, we wish to point out that 29.0 $J/\Psi \to \phi\pi^0\eta$ events in the resonance region
relative to 1000  $J/\Psi \to \phi\pi^+\pi^-$ in the same kinematic range
may look like a mere 3\% effect of isospin violation, but,
due to the lack of interference, it  corresponds to about 17\% isospin breaking on the amplitude level.
Therefore, $a_0-f_0$ mixing leads indeed to a very sizeable isospin-violating signal.

\begin{figure}
\includegraphics[width=0.925\linewidth]{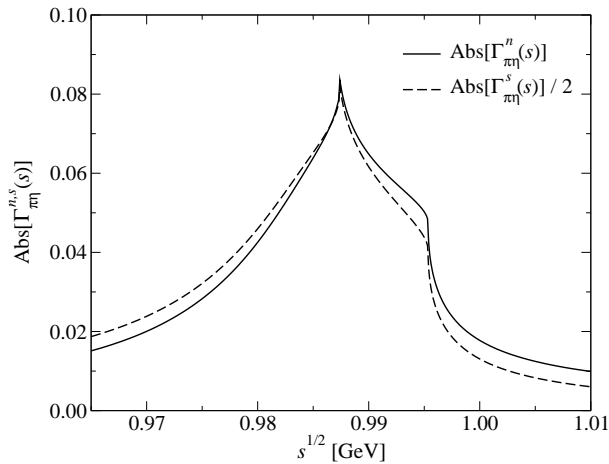}
\caption{\label{fig:AbsFF}
Absolute values of the scalar form factors
$\Gamma^n_{\pi\eta}(s)$ (full line), $\Gamma^s_{\pi\eta}(s)$ (dashed line).
Note that, to facilitate the comparison,
the latter has been scaled down by a factor of 2.
}\end{figure}

\begin{figure}
\includegraphics[width=0.925\linewidth]{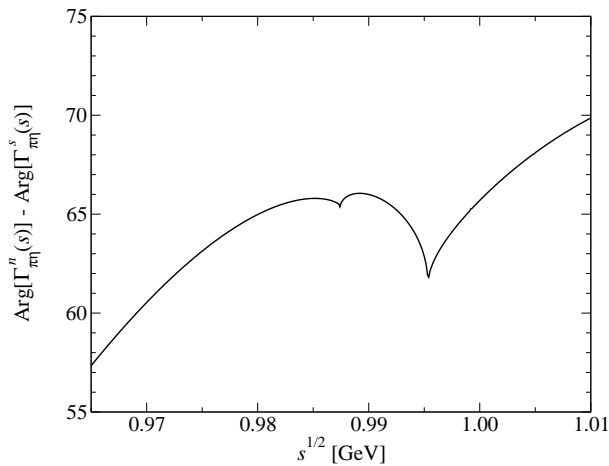}
\caption{\label{fig:PhaseFF}
Relative phase of non-strange and strange $\pi\eta$ scalar form factors,
${\rm Arg}\bigl[\Gamma^n_{\pi\eta}(s)\bigr]-{\rm Arg}\bigl[\Gamma^s_{\pi\eta}(s)\bigr]$.
}\end{figure}

As a side remark, we briefly comment on the form factors $\Gamma^n_{\pi\eta}(s)$,  
$\Gamma^s_{\pi\eta}(s)$ individually, which in principle may occur in a different
relative combination when $a_0-f_0$ mixing is investigated in the context of
a different decay or production mechanism.
In Figs.~\ref{fig:AbsFF} and \ref{fig:PhaseFF}, we show the absolute values of the form factors
as well as the \emph{difference} of the phase motions of the two 
(both including all isospin-breaking mechanisms discussed earlier).  
We find that the shape of (the absolute values of) both form factors is very similar, 
and that the phase difference, while showing remnants of the cusps, varies rather
mildly over the two-kaon threshold region (by altogether less than $15^\circ$).  
We conclude from these observations that, 
if the relative strength parameter $\lambda_\phi$ is replaced
by a different, possibly complex (but still approximately constant) value, we would not expect
the mixing signal to be wildly different from what we predict here.

\section{Summary}\label{sec:summary}

In this work we have improved the theoretical understanding of
the phenomenon of $a_0-f_0$ mixing. 
We have confirmed that the dominant mixing effect comes
from the kaon mass difference in line with Ref.~\cite{achasovmix}.
It is therefore possible to extract independent information
on the effective couplings of $a_0(980)$ and $f_0(980)$
to kaons, which
contain important structure information~\cite{evidence},
from the mixing matrix element. 

We have applied our formalism to the reaction $J/\Psi\to \phi \pi^0\eta$ and give
what we consider the best and most comprehensive prediction to date for this decay channel. 
The corresponding measurement will soon be possible with the upgraded BESIII detector~\cite{wu}.

In addition we addressed the following items:
\begin{enumerate}
\item As it is illustrated in Fig.~\ref{fig:events}, although the kaon
loop effect dominates the isospin-violating signal, the presence
of isospin-violating coupling constants introduces a significant additional
enhancement by roughly 50\% (corresponding to a 20\% effect on the
amplitude level). To the order we are working, 
isospin violation in the vertices and in the kaon masses are of the same
origin (i.e.\ are calculated from the same chiral Lagrangian), and no additional parameters enter.
\item The effect of soft photon exchange in the meson propagation is small,
even in the signal region, and amounts to a reduction of the signal
by 10\% (or 5\% on the amplitude level).
\item Neither of the two additional effects studied in this work, 
associated with an isospin-violating production operator,
distorts the shape of the signal severely,
c.f.\ gray band vs.\ solid line in Fig.~\ref{fig:isoProd}. 
\item We have confirmed that
the isospin violation in the final-state interaction, which can be identified with
$a_0-f_0$ mixing, is much more important than isospin violation in 
the production operator. 
We therefore confirm the corresponding claim of Refs.~\cite{aipproc,report}
for the particular case of the reaction $J/\Psi\to \phi \pi^0\eta$.
\end{enumerate}
Our final results are presented for  a particular 
reaction. However, the formalism to calculate the
propagating $a_0-f_0$ system in the presence of
isospin violation is very general and
therefore could also be used for the analysis
of other experiments, once data is available.
Especially items 1--3 of the list given above are
independent of the reaction studied,
and ought to be distinctly different from predictions of
models viewing $a_0$ and $f_0$ as $q\bar q$ or four-quark systems,
which allow for direct mixing without two-kaon-threshold enhancement.
We hence conclude that it is indeed possible
to measure and interpret $a_0-f_0$ mixing.

\begin{acknowledgments}

We wish to thank Timo L\"ahde for very helpful communications and e-mail exchanges,
and in particular for making preliminary fit results and numbers on the
three-channel generalization of Ref.~\cite{timo} available to us.
B.K.\ and J.R.P.\ are grateful for the hospitality of 
the Institut f\"ur Kernphysik at the Forschungszentrum J\"ulich 
in the initial stage, and 
B.K.\ for that of the Facultad de Ciencias F\'isicas at the 
Universidad Complutense Madrid in the final stage of this project.
The authors acknowledge partial financial support by the 
EU I3HP Project (RII3-CT-2004-506078).
B.K.\ was supported by the DFG (SFB/TR 16).
The research of J.R.P.\ was partially funded by Spanish CICYT contracts
FPA2005-02327 and FIS2006-03438, as well as Banco Santander/Complutense
contract PR27/05-13955-BSCH.
\end{acknowledgments}

\appendix

\section{Charge vs.\ isospin basis}\label{app:basis}

The two-meson states of definite isospin (in the $S$-wave) are related
to the states in the charge basis by the following 
relations:
\begin{widetext}
\beq
  \label{eq:changeofbasis}
  \left(
    \begin{array}{c}
|\pi\pi\rangle_{I=2}\\
|\pi\pi\rangle_{I=0}\\
|\eta\eta\rangle_{I=0}\\
|K\bar K\rangle_{I=0}\\
|K\bar K\rangle_{I=1}\\
|\pi^0\eta\rangle_{I=1}\\
    \end{array}
\right)_{J=0}
\equiv
\left(\begin{array}{ccccccc}
-\sqrt{1/6}&\sqrt{1/3}&0&0&0&0\\
-\sqrt{1/3}&-\sqrt{1/6}&0&0&0&0\\
0&0&1/\sqrt{2}&0&0&0\\
0&0&0&-\sqrt{1/2}&-\sqrt{1/2}&0\\
0&0&0&-\sqrt{1/2}&\sqrt{1/2}&0\\
0&0&0&0&0&1\\
\end{array}\right)
  \left(
    \begin{array}{c}
|\pi^+\pi^-\rangle\\
|\pi^0\pi^0\rangle\\
|\eta\eta\rangle\\
|K^+ K^-\rangle\\
|K^0 \bar K^0\rangle\\
|\pi^0\eta\rangle\\
    \end{array}
\right)_{J=0} ~,
\end{equation}
\end{widetext}
where $|\pi^+\pi^-\rangle$ denotes the symmetrized combination of $|\pi^+\pi^-\rangle$ 
and $|\pi^-\pi^+\rangle$.  The additional factors of $1/\sqrt{2}$ for the $\pi\pi$
and the $\eta\eta$ states account for Bose symmetry.
Using the isospin basis for our calculation is particularly convenient 
as we consider production of states with definite isospin ($I=0$ mostly), 
and the final state $\pi^0\eta$ has definite isospin $I=1$, too.

\section{The loop function \boldmath{$G$}}\label{app:G}

For two mesons $A$, $B$  with masses $M_A$ and $M_B$ propagating in the loop, 
the elementary two-point function is given by
\beq \begin{split}  \label{eq:gdifferentmasses}
G_{AB}(s) &= -\frac{1}{16\pi^2}\Biggl\{
\frac{\Delta}{s} \biggl[
\log\frac{1+\Omega_A}{1+\Omega_B} -\log\frac{M_A}{M_B}  \biggr] \\
&+\frac{\nu}{2s} 
\log\frac{(\nu\Omega_A+s-\Delta)(\nu\Omega_B+s+\Delta)}
{(\nu\Omega_A-s+\Delta)(\nu\Omega_B-s-\Delta)} \\
& -\log\biggl[ \frac{q^2_{max}}{M_A M_B}  \bigl(1+\Omega_A\bigr)\bigl(1+\Omega_B\bigr)\biggr] 
\Biggr\} ~,
\end{split} \eeq
where $\Delta=M_B^2-M_A^2$, $\nu^2=s^2-2 s (M_A^2+M_B^2)+\Delta^2$, 
$\Omega_{A,B}=\sqrt{1+M_{A,B}^2/q_{\max}^2}$\,,
and $q_{max}$ is the cutoff.
The form in Eq.~\eqref{eq:gdifferentmasses} is correct and unambiguous below
the pseudothreshold, $s \leq (M_A-M_B)^2$, and has to be continued analytically
into other kinematical regions.
As described in the text, the determination of $G_i(s)$ has to be chosen so that on the physical, 
right hand, cut of the $i$-th state,
$\im G_i(s)=\sigma_i(s)/16\pi=k_i/(8\pi\sqrt{s})$,
where $k_i$ stands for the corresponding center-of-mass momentum.
In the equal-mass limit $M_A=M_B=M$, Eq.~\eqref{eq:gdifferentmasses} reduces to $G_{A=B}(s)=G(s)$, 
\beq  \label{eq:gsamemass}
  G(s) =-\frac{1}{16\pi^2}\biggl\{
\sigma\log\frac{\sigma\Omega+1}{\sigma\Omega-1} 
-2\log\biggl[\frac{q_{max}}{M}(1+\Omega)\biggr]
\biggr\} ~,
\eeq
where $\sigma=\sqrt{1-4M^2/s}$\,, $\Omega=\sqrt{1+M^2/q_{\max}^2}$\,.

\section{Polynomial remainders of the scalar form factors}\label{app:R}

The polynomial remainders of the scalar form factors as defined in 
Eq.~\eqref{eq:Rdef},
up to $\Order(p^4)$ in ChPT, are given as
\beq \begin{split}
R_\pi^n(s) &= \sqrt{\frac{3}{2}} \biggl\{ 1 +
\mu_{\pi(\pi)} - \frac{\mu_{\eta(\pi)}}{3} +
\frac{16\mpi}{\fp^2}        \bigl(2L_8^r-L_5^r\bigr) \\ &+
\frac{8(2\mk+3\mpi)}{\fp^2} \bigl(2L_6^r-\!L_4^r\bigr) +
\frac{4s}{\fp^2}            \bigl(2L_4^r+L_5^r\bigr) 
\biggr\} \,,\\
R_\pi^s(s) &= \frac{\sqrt{3}}{2} \biggl\{ 
\frac{16\mpi}{\fp^2} \bigl(2L_6^r-L_4^r\bigr) +  
\frac{8s}{\fp^2}            L_4^r
\biggr\} \,,\\
R_K^n(s) &= \frac{1}{\sqrt{2}} \biggl\{ 1 +
\frac{2}{3}\mu_{\eta (K)} +
\frac{16\mk}{\fk^2}        \bigl(2L_8^r-L_5^r\bigr) \\ &+
\frac{8(6\mk+\mpi)}{\fk^2} \bigl(2L_6^r-L_4^r\bigr) +
\frac{4s}{\fk^2}           \bigl(4L_4^r+L_5^r\bigr) 
\biggr\} \,,\\
R_K^s(s) &= 1 +
\frac{2}{3}\mu_{\eta (K)}+ 
\frac{16\mk}{\fk^2}        \bigl(2L_8^r-L_5^r\bigr) \\ &+
\frac{8(4\mk+\mpi)}{\fk^2} \bigl(2L_6^r-L_4^r\bigr) +
\frac{4s}{\fk^2}           \bigl(2L_4^r+L_5^r\bigr) 
\,,  \\
R_\eta^n(s) &= - \frac{1}{3\sqrt{2}} \biggl\{ 1 -
3\mu_{\pi(\eta)}+4\mu_{K(\eta)}-\frac{\mu_{\eta(\eta)}}{3} \\ &+ 
\frac{16\me}{\fe^2}          \bigl(2L_8^r-\!L_5^r\bigr) +
\frac{8(10\mk-\mpi)}{\fe^2}  \bigl(2L_6^r-L_4^r\bigr) \\ &-
\frac{128(\mk-\mpi)}{3\fe^2} \bigl(3L_7+\!L_8^r\bigr)  +
\frac{4s}{\fe^2}             \bigl(6L_4^r+L_5^r\bigr) 
\biggr\} \,,\\
R_\eta^s(s) &= -\frac{2}{3} \biggl\{ 1 +
2\mu_{K(\eta)} -\frac{4}{3}\mu_{\eta(\eta)} \\ &+ 
\frac{16\me}{\fe^2}        \bigl(2L_8^r-L_5^r\bigr) +
\frac{4(8\mk+\mpi)}{\fe^2} \bigl(2L_6^r-L_4^r\bigr) \\ &-
\frac{64(\mk-\mpi)}{3\fe^2} \bigl(3L_7+\!L_8^r\bigr) +
\frac{2s}{\fe^2}           \bigl(3L_4^r+2L_5^r\bigr) 
\biggr\} \,,
\end{split}\eeq
where $\mu_{i(j)}$ are the tadpole loop functions defined as
\beq
\mu_{i(j)} \eq \frac{M_i^2}{32\pi^2F_i F_j}\log\frac{M_i^2}{\mu^2} ~,
\eeq
and this particular choice of decay constants $F_{i/j}$ is in accordance
with Ref.~\cite{TimoPriv}.
The $L_i^r$ are the standard low-energy constants defined in Ref.~\cite{GL85}.

\section{Scattering amplitudes with isospin violation}\label{app:scattamps}

In this appendix we give a complete list of $S$-wave projected 
four-meson amplitudes  derived from Eq.~\eqref{L2} for
those channels that are charge- and strangeness-neutral, i.e.\ 
reactions linking the channels (1) $\pi^+\pi^-$, (2) $\pi^0\pi^0$, (3) $\eta\eta$,
(4) $K^+K^-$, (5) $K^0\bar{K}^0$, (6) $\pi^0\eta$ including isospin violation;
see also Ref.~\cite{Beisert}.
Our normalization for the $S$-wave projection is given by
\beq
T_{ab}^{J=0}(s) = \frac{1}{2}\int_{-1}^1 dz\, T_{ab}\bigl(s,t(s,z),u(s,z)\bigr) ~. 
\eeq
$a$, $b$ refer to the numbering of the possible in- and out-channels 1--6 as given above;
of course one has $T_{ab}=T_{ba}$.
In the following list, for reasons of brevity we suppress the $J=0$ superscripts:
\begin{align*}
T_{11} &\eq 2 T_{14} \eq T_{44} \eq 
\frac{s+4\Dpi}{2F^2}  ~,\\
T_{12} &\eq 
\frac{s-\mpin}{F^2} ~,\\
T_{13} &\eq \frac{1}{3}T_{22} \eq T_{23} \eq T_{66} \eq
\frac{\mpin}{3F^2}  ~, \\
T_{15} &\eq  T_{45} \eq \frac{1}{2}T_{55} \eq
\frac{s}{4F^2} ~,\\
T_{16} &\eq  
\frac{\eps}{3F^2} \bigl(4\mpi-3s\bigr) ~,\\
T_{24} &\eq   T_{25} \eq 
\frac{s}{4F^2}\bigl(1\pm2\sqrt{3}\eps\bigr) \mp\frac{2\eps}{\sqrt{3}F^2}\mk ~,\\
T_{26} &\eq  - T_{36} \eq 
-\frac{4\eps}{3F^2}\bigl(\mk-\mpi\bigr) ~,\\
T_{33} &\eq   
\frac{4\me-\mpin}{3F^2}~,\\
T_{34} &\eq 
\frac{3s}{4F^2}\Bigl(1-\frac{2\eps}{\sqrt{3}}\Bigr) 
-\frac{2}{3F^2}\bigl(\mkc-\Dpi\bigr)\bigl(1-\sqrt{3}\eps\bigr)  ~,
\end{align*}
\begin{align}
T_{35} &\eq 
\frac{3s}{4F^2}\Bigl(1+\frac{2\eps}{\sqrt{3}}\Bigr) 
-\frac{2\mkn}{3F^2}\bigl(1+\sqrt{3}\eps\bigr)    ~,\no\\
T_{46} &\eq  
\frac{1}{4\sqrt{3}F^2}\Bigl(1+\frac{2\eps}{\sqrt{3}}\Bigr) 
\Bigl(3s-4\bigl(\mkc-\Dpi\bigr)\Bigr) ~,\no\\
T_{56} &\eq 
-\frac{1}{4\sqrt{3}F^2}\Bigl(1-\frac{2\eps}{\sqrt{3}}\Bigr)
\bigl(3s-4\mkn\bigr)  ~.
\end{align}
All amplitudes are normalized by a factor of $1/F^2$ which, at this
accuracy, can arbitrarily be identified with any meson decay constant.
For numerical evaluations  
we use the convention that the overall $1/F^2$ factor is replaced 
by one $1/\sqrt{F_\phi}$ factor for every
external meson $\phi$ in the process concerned; for a discussion
on how and why this choice yields the best description of data
see Refs.~\cite{Pelaez:2004xp,Descotes-Genon:2003cg}. 

\medskip

\section{Numerical input}\label{app:numbers}

In our calculations we use the masses
$M_{\pi^+}=139.57\,$MeV, $M_{\pi^0}=134.98\,$MeV, 
$M_{K^+}=493.68\,$MeV, $M_{K^0}=497.67\,$MeV, 
$M_\eta=547.8\,$MeV,
$M_{J/\Psi} = 3097\,$MeV, $M_\phi = 1020\,$MeV.
In addition we need the decay constants $\fp=92.4\,$MeV,
$\fk=1.22 \fp$ and $\fe=1.3 \fp$.
The leading order $\pi^0\eta$ mixing angle is $\eps = 0.01$.
For the low-energy constants of order $p^4$~\cite{GL85} needed for the
polynomial terms of the scalar form factors, we use the 
numerical values
$L_4^r= 0.84\times 10^{-3}$, 
$L_5^r= 0.52\times 10^{-3}$, 
$L_6^r=-0.18\times 10^{-3}$, 
$L_7  =-0.40\times 10^{-3}$, 
$L_8^r= 0.15\times 10^{-3}$
(all given at a scale $\mu=M_\rho$), 
as obtained in a three-channel generalization~\cite{TimoPriv}
of the formalism presented in Ref.~\cite{timo}.
From the same fit, we use the relative strength parameter
of non-strange to strange scalar source terms,
$\lambda_\phi = 0.117$.
In the loop function $G(s)$ (see Appendix~\ref{app:G}),
we employ a cutoff $q_{max}=0.95\,$GeV.

\end{document}